\documentclass[%
article, two column,
superscriptaddress,
preprintnumbers,
 amsmath,amssymb,
 aps,
 prd,
longbibliography.usenames, dvipsnames]{revtex4-2}
\usepackage[main=english]{babel}
\usepackage{quantikz}
\usepackage{amssymb}
\usepackage{graphicx} 
\usepackage[utf8]{inputenc}

\usepackage{subfig}

\usepackage{dcolumn}
\usepackage{siunitx}
\usepackage{dsfont}

\usepackage[utf8]{inputenc}
\usepackage[toc,page]{appendix}

\usepackage{ragged2e}
\usepackage{graphicx}
\usepackage{hyperref}
\hypersetup{colorlinks}
\usepackage{soul}

\usepackage{physics}
\usepackage{bbold}
\usepackage{float}
\usepackage{hyphenat}

\definecolor{azure(colorwheel)}{rgb}{0.0, 0.5, 1.0}
 \definecolor{babyblue}{rgb}{0.54, 0.81, 0.94}
\definecolor{cornflowerblue}{rgb}{0.39, 0.58, 0.93}

\usepackage[justification=justified]{caption}

\usepackage{amssymb}

\newcommand{\pushright}[1]{\ifmeasuring@#1\else\omit\hfill$\displaystyle#1$\fi\ignorespaces}

\begin{document}

\hypersetup{colorlinks=true,citecolor=azure(colorwheel),linkcolor=magenta,filecolor=magenta,urlcolor=azure(colorwheel),breaklinks=true}

\newcommand{\TE}[1]{\mathcal{U}(#1)}

\author{Oriel~Kiss}
\email{orielkiss@gmail.com}
\affiliation{Department of Nuclear and Particle Physics, University of Geneva, Geneva 1211, Switzerland}
\affiliation{European Organization for Nuclear Research (CERN), Geneva 1211, Switzerland}

\affiliation{Physics Department, University of Trento, Via Sommarive 14, I-38123 Trento, Italy}

\author{Ivano~Tavernelli}
\affiliation{IBM Quantum, IBM Research Europe – Zurich, 8803 Rueschlikon, Switzerland}
\author{Francesco~Tacchino} 
\affiliation{IBM Quantum, IBM Research Europe – Zurich, 8803 Rueschlikon, Switzerland}

\author{Denis Lacroix}
\affiliation{Université Paris-Saclay, CNRS/IN2P3, IJCLab, 91405 Orsay, France}

\author{Alessandro~Roggero}
\affiliation{Physics Department, University of Trento, Via Sommarive 14, I-38123 Trento, Italy}
\affiliation{INFN-TIFPA Trento Institute of Fundamental Physics and Applications, Trento, Italy}

\title{Neutrino thermalization via randomization on a quantum processor}

\begin{abstract}
The dynamical evolution of neutrino flavor in supernovae can be modeled by an all-to-all spin Hamiltonian with random couplings. Simulating such two-local Hamiltonian dynamics remains a major challenge, as methods with controllable accuracy require circuit depths that increase at least linearly with system size, exceeding the capabilities of current quantum devices. The eigenstate thermalization hypothesis predicts that these systems should thermalize, a behavior confirmed in small-scale classical simulations. Here we investigate flavor thermalization in much larger systems using random quantum circuits as an empirical tool to emulate the non-local dynamics, and demonstrate that thermal behavior can be reproduced using a depth independent of system size. By simulating systems of over one hundred qubits, we find that the thermalization time grows approximately as the square root of the system size, consistent with predictions from semi-classical methods. Our study also illustrates that near-term quantum devices are useful tools to test and validate empirical classical methods, and highlights an application of random circuits in physics.

\end{abstract}

\date{\today}
\maketitle

\section{Introduction}

In astrophysical scenarios marked by high temperature and density, such as core collapse supernovae and binary neutron star mergers, the emission of neutrinos in substantial fluxes plays a pivotal role in the dynamical evolution of the system and the nucleosythesis that takes place in these environments~\cite{pantaleone1992neutrino,Janka:2006fh,Woosley:2005,Hoffman:1997}. Neutrinos interact through weak currents with local nucleons, influencing the proton-to-neutron ratio.  
A comprehensive understanding of neutrino flavor content is essential for detailed investigations into the evolution of these systems \cite{Sigl:1992fn,pastor2002flavor,pastor2002physics,Bell:2003mg,Balantekin_2007}. In dense regions, neutrinos can experience coherent forward scattering, a process highly sensitive to their quantum mechanical flavor states. Flavor-dependent relative phases may arise during scattering on charged leptons, leading to the well-known MSW effect, but they can also be generated through neutral current interactions with other neutrinos in the system~\cite{pantaleone1992neutrino,pastor2002flavor,Raffelt:2007xt}. In scenarios with a sufficiently high neutrino number density, this flavor exchange effect is expected to dominate the neutrino gas's flavor evolution, giving rise to  coherent phenomena like flavor spectrum splits and swaps~\cite{Duan:2005cp,Duan:2006jv,Duan2010review}. {The phenomenology of collective neutrino oscillations has been traditionally explored by means of mean-field methods (see~\cite{Duan2010review,Chakraborty_2016,annurev:/content/journals/10.1146/annurev-nucl-121423-100853} for reviews) but in recent years a renewed interest in the role played by quantum correlations have emerged (see~\cite{Patwardhan2020,Balantekin_2023,Volpe_review} for recent reviews).}

In this paper, building on Ref.~\cite{Equilibirum_neutrio_roggero}, we focus on investigating the flavor evolution of a neutrino gas under the condition where the potential arising from coherent $\nu-\nu$ forward scattering prevails as the dominant effect. This scenario is commonly referred to in the literature as the ``fast'' flavor oscillation limit \cite{sawyer2004classical,chakraborty2016self,dasgupta2017fast,Martin:2019dof,Roggero2022,Xiong:2021dex}. In this limit, the redistribution of flavor among neutrinos is anticipated to transpire over length scales of approximately one meter. Over these length scales we assume the changes in the interaction potential, due to both the neutrino density becoming smaller as we move away from the emission region as well as neutrinos becoming progressively more forward peaked, to be negligible. Our exploration is centered on this specific regime, aiming to comprehensively analyze the associated dynamics. We study this regime using randomized quantum simulations strategies based on product formula, previously tackled using standard Trotter Suzuki decomposition by Refs~\cite{PRD_neutrino,neutrino_simulation_amitrano}. In this first study we employ the common approximation where only two flavors are considered for each neutrino, this is motivated by the expectation that in a core collapse supernova the number density and fluxes of heavy flavor neutrinos are comparable since they decouple in the same region. Genuine multi-flavor effects might be important in some regimes (see e.g.~\cite{PhysRevLett.125.251801,PhysRevD.103.063013}) and extensions of our simulation strategy to the full three-flavor case, and the inclusion of anti-neutrino, could be an avenue for future research.

We carry out our computational investigations by leveraging, besides classical benchmarks, the opportunities offered by state-of-the-art digital quantum simulators~\cite{simulation_tacchino,miessen2023quantum}. In recent years, this technology has advanced significantly, enabling so-called ``quantum utility'' experiments~\cite{kim2023evidence,fischer2024dynamical}, in which noisy quantum processors augmented by error mitigation techniques successfully executed quantum circuits beyond the reach of exact classical simulation. These pioneering demonstrations marked the initial steps toward employing quantum computers as research tools in fundamental physics, which can be used to validate theoretical predictions and eventually to explore regimes previously inaccessible to classical methods. Several utility-scale experiments have since been reported in high-energy physics~\cite{dimeglio2024quantum,halimeh2025quantum}, with a primary focus on simulating scattering dynamics~\cite{Chai2025fermionicwavepacket,schuhmacher2025observation,chai2025resource,Roland_2024,zemlevskiy2024,farrell2025} and in general on lattice gauge theories~\cite{Farrell2024scalable,cobos2025real,cochran2025visualizing}. In this work, we introduce quantum simulation techniques into the domain of neutrino physics, presenting results at the 100-qubit scale.

We study neutrino thermalization, modeled as an all-to-all spin Hamiltonian with random couplings, tracking thermalization through the flavor variance observable. To establish a benchmark, we first perform small-scale simulations using a deterministic Trotterized product formula with controlled accuracy. We then introduce a randomized product formula that reproduces the non-local dynamics with circuit depth independent of system size, along with its theoretical justification. Deploying this approach classically and on IBM superconducting quantum hardware for systems of up to 112 neutrinos, we find that the thermalization time grows as the square root of the system size across both backends. This consistent agreement validates semi-classical treatments of collective neutrino oscillations and demonstrates that near-term quantum devices can serve as independent benchmarks for classical approximation methods.

\section{Methods}
\subsection{Model and thermalization}
\label{sec:model}
The $\nu-\nu$ coherent forward scattering potential takes the form of an all-to-all Heisenberg-like interaction~\cite{Balantekin_2007,Pehlivan2011,Sonner2017,harrow2022thermalization}, and such Hamiltonians are generically expected to be nonintegrable except in special cases~\cite{Equilibirum_neutrio_roggero,PhysRevD.107.043024,PhysRevResearch.7.023157,neill2025universalitysuinftyrelaxationdynamics,PhysRevC.110.045801}. Furthermore, non-integrable Hamiltonians are expected to “thermalize” \cite{Srednicki_1996,Deutsch91} in the sense that expectation values of few-body operators tend to equilibrate to a value that can be predicted from an appropriate thermodynamic partition function (see~\cite{D'Alessio03052016} for a review). Recent work has also demonstrated thermalization induced by random all-to-all interactions on a quantum processor~\cite{perrin2025dynamic}.
For the rest of this work, we assume that the vacuum oscillation potential is negligibly small relative to the other two potentials. We also assume that the matter profile is uniform in the regions of the environment under consideration, allowing us to consider a corotating frame to remove its overall effect on each neutrino. After these modifications, only the $\nu-\nu$ coherent forward scattering Hamiltonian remains. We assume axial symmetry for the velocity coupling term, enforcing that the velocity components orthogonal to the momentum symmetry axis (denoted by $z$) average to zero. This leads to the following Hamiltonian for $N$ neutrinos~\cite{Balantekin_2007,Pehlivan2011}
\begin{equation}
\begin{split}
\hat{H}_{\nu\nu} &= \frac{\mu}{2N}\sum_{i<j} (1 - {\bf v}_i \cdot {\bf v}_j)\hat{\vec{\sigma}}_i \cdot \hat{\vec{\sigma}}_j \\
&\equiv H_0 +  \frac{\mu}{N} \sum_{i<j} (1 - {\bf v}_i \cdot {\bf v}_j) P_{i,j}. 
\end{split}
\label{eq:nu_ham}
\end{equation}
Here, ${\bf v}_k$ is a three-dimensional real vector containing the velocity of the $k$-th neutrino, $\hat{\vec{\sigma}}$ is a vector of Pauli operators, $\mu$ is the scale of the $\nu-\nu$ interaction, $P_{i,j}$ is a swap operation which can be expressed as 
\begin{equation}
    P_{i,j} = \frac{1}{2}(\mathbb{1} + \vec{\sigma}_i\cdot\vec{\sigma}_j),
    \end{equation}  
and $H_0$ an irrelevant constant. Given a typical core-collapse supernova flux at a radius of 50 km, a simple estimate gives $\mu \approx 1 \, \text{cm}^{-1}$. As it is the only dimensional parameter in the problem, we hereafter measure all distances and times in units of $\mu$ and set $\mu = 1$. We consider a forward peak distribution with $v_z \sim 1- \frac{2}{\sqrt{\pi}\sigma }\Theta(x)e^{-x^2/2\sigma^2}$ with $\sigma=0.5$. The remaining coordinates are then fixed as 
\begin{equation}
\begin{split}
\label{eq:dist_vel}
    v_{x}  &= \sqrt{1-v_z^2} \cos{\theta} \\
    \text{and }  v_{y}  &= \sqrt{1-v_z^2} \sin{\theta} \\
      \text{where }\theta &\sim U(0,2\pi).
\end{split}
\end{equation}
This choices for modeling the angular distribution of neutrinos were previously used also in Ref.~\cite{Equilibirum_neutrio_roggero} where the connection between the chaotic nature of the Hamiltonian and flavor thermalization was investigated for small systems with $N\leq16$. The rationale behind the use of a random distribution for the velocities is twofold: on one hand it captures the expectation for the geometry of a core collapse supernova with mostly outgoing neutrinos and, at the same time, allows to avoid possible spurious symmetries that can be introduced in the calculation when using uniform grids of angles~\cite{Martin:2023ljq,Equilibirum_neutrio_roggero}.

In this work we are interested in the typical dynamics generated by the neutrino interactions and, given the presence of random couplings, we will consider an ensemble of Hamiltonians $\{H^{(k)}\}_{k=1,\dots,N_H}$ defined explicitly as follows
\begin{equation}
\label{eq:Hamiltonian_sample}
\hat{H}_{\nu\nu}^{(k)} = \frac{\mu}{N} \sum_{i<j} (1 - {\bf v}^{(k)}_i \cdot {\bf v}^{(k)}_j) P_{i,j} \equiv \sum_{i<j} \mu^{(k)}_{i,j} P_{i,j}\;,
\end{equation}
with the associated unitary time evolution denoted as
\begin{equation}
\mathcal{U}^{(k)}(t) = \exp\left(-it\hat{H}_{\nu\nu}^{(k)}\right)\;.
\end{equation}
Previous work \cite{Equilibirum_neutrio_roggero} estimated the thermalization time $T_{\text{th}}$, or at least a lower bound, using single-qubit observables such as the flavor expectation value for a product state $|\psi_0\rangle$:
\begin{equation}
Z^{(k)}_j(t) = \langle \psi_0|\mathcal{U}^{(k)}(t)^\dagger Z_j \mathcal{U}^{(k)}(t)| \psi_0 \rangle.
\end{equation}
A characteristic time was defined as the first inversion point, i.e the first local minimum or maximum. Additional estimates were obtained from the time when the single-qubit entropy reached $95\%$ of its maximum, or from global measures such as the Loschmidt echo,
\begin{equation}
L^{(k)}(t) = |\langle \psi_0|\mathcal{U}^{(k)}(t)|\psi_0 \rangle|^2,
\end{equation}
with $T_{\text{th}}$ identified from the first minimum of $L^{(k)}(t)$.

In contrast, the randomization approach used here does not track individual neutrinos, but instead allows us to evaluate observables invariant under qubit relabeling. We therefore introduce the flavor variance,
\begin{equation} \label{Eq_V_k_t}
V^{(k)}(t)= \frac{1}{N}\sum_j Z^{(k)}_j(t)^2\;. 
\end{equation}
Since every neutrino thermalizes to the same asymptotic state, the variance vanishes at long times. As a proxy for the typical time scale, we define $T_{\text{th}}$ as the first time $V^{(k)}(t)$ drops below $\Gamma = 0.1$.
\begin{figure}
    \centering
    \includegraphics[width=0.45\textwidth]{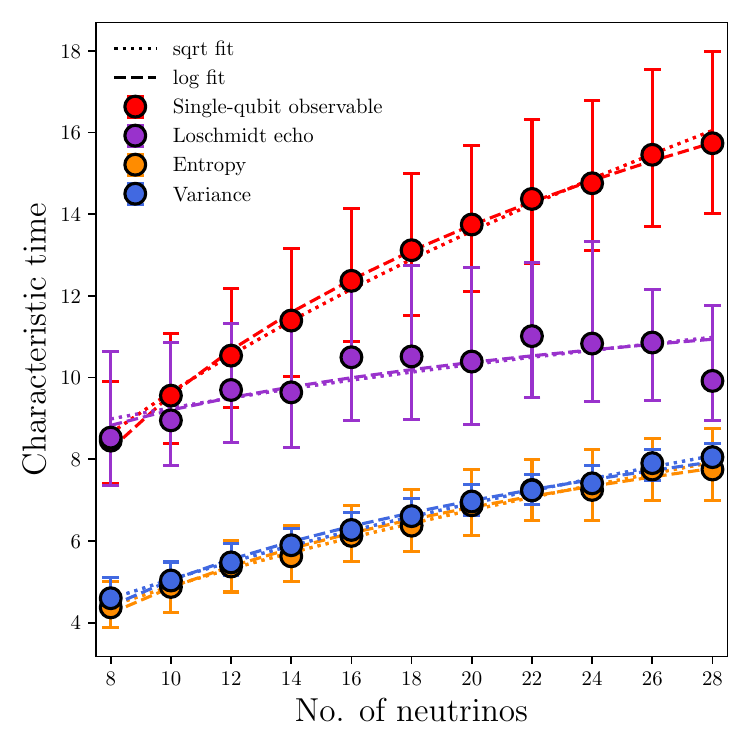}
    \caption{\justifying\textbf{Trotterized dynamics.} Scaling of the characteristic time as function of the system size for the single qubit observables (red), Loschmidt echo (purple), single qubit entropy (orange) and variance (blue), alongside square root and logarithm fits (dotted and dashed lines). The calculation are performed with a converged second-order Trotter product formula ($\delta t = 0.5$). Error bars correspond to a 25\%-75\% confidence interval over random Hamiltonian realizations.}
    \label{fig:trotter}
\end{figure}

Fig.~\ref{fig:trotter} shows the characteristic time of the single-qubit observable, the entropy, as well as the variance and Loschmidt echo, where the evolution is nearly exact at using a converged Trotter decomposition with time step $\delta t = 0.5$, acting on the domain-wall configuration, $|\phi_0\rangle = |0\rangle^{\otimes N/2} \otimes |1\rangle^{\otimes N/2}$. The errorbars represent a 10\%–90\% confidence interval over different Hamiltonian realizations.

We first observe that the variance and entropy exhibit consistent behavior, which is significant since our random quantum channel only allows access to permutation-invariant observables. Therefore, the variance is a reasonable observable to look for thermalization, since it is consistent with the predictions using the entropy. Second, in the absence of simulation of larger systems, the current data is insufficient to distinguish between a square-root and a logarithmic scaling; both remain statistically compatible. Consequently, simulations at larger system sizes are required to resolve this ambiguity. 
Below we discuss several method to treat approximately the problem on a Noisy Intermediate Scale Quantum (NISQ) devices \cite{Preskill2018quantumcomputingin}, such as Trotterization and our randomized channel. We do not delve into asymptotical optimal methods such as Quantum Singular Value Transformation \cite{GilyenExponentialImprovement} as the overhead is too high for current platforms.

\subsection{Trotterized dynamics}
\label{sec:trotter}
 As was shown in previous works~\cite{PRD_neutrino,neutrino_simulation_amitrano} a Trotter decomposition of the evolution operator can be implemented on a qubit chain by using a SWAP network where every pair interaction is followed by a SWAP. With this strategy a full Trotter step requires $N/2$ layers of nearest neighbor interactions. In order to express this construction in a compact way, let us first define a single layer unitary which applies nearest neighbor Heisenberg interactions on the $N/2$ even pairs followed by the $N/2-1$ odd pairs
\begin{equation}
\begin{split}
W(t,{\bf h},{\bf p}) = &\left[\prod_{j\;even} e^{-ith_{j} P_{j,j+1}}P^{p_j}_{j,j+1}\right]\\
&\times\left[\prod_{j\;odd} e^{-ith_{j} P_{j,j+1}}P^{p_j}_{j,j+1}\right]\;,
\end{split}
\label{eq:hamden}
\end{equation}
where the vector ${\bf p}$ has $N-1$ components that can be either zero or one: if $p_j=1$ then we perform a SWAP operation between qubits $j$ and $j+1$ while we implement the interaction between them. Instead, the $N-1$ component vector ${\bf h}$ is used to retrieve the correct coupling constants for interactions. Using the fact that $P_{i,j}$ is idempotent, a more convenient expression for $W(t,{\bf h},{\bf p})$ can be found to be
\begin{equation}
\begin{split}
\label{eq:wunitary}
W(t,{\bf h},{\bf p})= &\left[\prod_{j\;even} e^{-i\left(th_{j}+p_j\frac{\pi}{2}\right) P_{j,j+1}}\right]\\
&\times\left[\prod_{j\;odd} e^{-i\left(th_{j}+p_j\frac{\pi}{2}\right) P_{j,j+1}}\right]\;,
\end{split}
\end{equation}
which now makes it apparent that the addition of possible SWAP operations does not need to incur in additional gates being implemented~\cite{PhysRevD.111.043038}. Using this layer unitary as building block, and for an even number of particles $N$, a single first order Trotter step for the Hamiltonian in Eq.~\eqref{eq:Hamiltonian_sample} can then be written as
\begin{equation}
\label{eq:trotter}
\mathcal{U}^{(k)}_T(t) = \prod_{l=1}^{N/2} W(t,{\bf h}^{(k,l)},{\bf 1})\;,
\end{equation}
where ${\bf 1}$ is a $N-1$ component vector containing $1$ in every entry and the coupling constant vectors are given by
\begin{equation}
h^{(k,l)}_j = \frac{\mu}{N}\left(1-{\bf v}^{(k)}_{i(j,l)}\cdot{\bf v}^{(k)}_{ip(j,l)}\right) = \mu^{(k)}_{i(j,l),ip(j,l)}\;.
\end{equation}
The label $i(j,l)$ denotes the label of the neutrino associated to qubit $j$ at the $l-th$ layer and similarly $ip(j,l)$ denotes the label of the neutrino associated to qubit $j+1$ at the same layer. The mapping between $i(j,l)$ and the qubit $j$ depends on the permutations that have been used in previous layers, taking into account that for odd $j$ we have applied already the even qubit layer in $W$. For instance, when $N=4$ and we perform a SWAP at every step, i.e. $\vec{p}^{(l)}=\vec{1}$, we have
\begin{equation}
\begin{split}
i(0,1) = 0\quad& ip(0,1) = 1\quad i(2,1) = 2\quad ip(2,1) = 3\\
&i(1,1) = 0\quad ip(1,1) = 3\\
i(0,2) = 1\quad& ip(0,2) = 3\quad i(2,2) = 0\quad ip(2,2) = 2\\
&i(1,2) = 1\quad ip(1,2) = 2.\\
\end{split}
\end{equation}
The Trotter unitary can then be used to approximate the correct time evolution operator leading to
\begin{equation}
\left\|\mathcal{U}^{(k)}(t)-\mathcal{U}^{(k)}_T(t/M)^M\right\|\leq \frac{t^2}{M}C^{(k)}\;,
\end{equation}
where $C^{(k)}$ is a sum of norms of commutators~\cite{neutrino_simulation_amitrano}, and $M$ is the number of so-called Trotter steps.

\color{black}

\subsection{Randomized dynamics}
\label{sec:rnd_dyn}
Simulating the flavor dynamics of neutrinos using product formulas is challenging on current and near-term quantum hardware due to the required depth growing at least linearly with system size~\cite{neutrino_simulation_amitrano}. In recent works, interacting neutrino systems have been shown to display intricate many-body correlations (see e.g.~\cite{Roggero2021a,Roggero2022,PhysRevLett.130.221003,Martin:2023ljq,Lacroix:2022krq,Siwach:2022xhx,PhysRevResearch.7.023228}) from a phenomenological point of view however the most important physical observables are single-neutrino properties (these dominate reactions in the supernova and are those measured on Earth). Our goal in this work is to design an approximate method that is able to track the evolution of single neutrino (qubit) observables in an efficient manner. We shall think of our method as a physically motivated ansatz, which is empirically able to capture permutation-invariant observables using only a circuit depth of $\mathcal{O}(t)$ \textemdash independent of the system size\textemdash at least up to intermediate time, which we mean as a timescale sufficiently large that it cannot be accurately captured by a low-order Taylor expansion. Although rigorous guarantees are beyond the scope of this work, establishing them remains an important avenue for future research.

From the discussion on the Trotterized dynamics it is apparent that the full evolution circuit is very similar to the evolution of a, time-dependent, random Heisenberg chain. Correlations among the, otherwise random, rotation angles are critical to allow to uniquely associate a specific neutrino to an individual qubit at any given point in the various layers. The coupling constants $\mu^{(k)}_{i,j}$ in the Hamiltonian are random and uncorrelated with the neutrino labels $(i,j)$. Consequently, when focusing on a single qubit, the evolution of its density matrix depends solely on the number of partial SWAP operations it has undergone since the initial state at $t = 0$. However, because these partial SWAPs only involve neutrinos with misaligned spins, the resulting qubit-level spin dynamics will generally depend on the initial assignment of qubits to neutrinos, even when the total flavor (i.e., the spin magnetization) is conserved. For instance, if we consider two states
\begin{equation}
\begin{split}
\label{eq:init_states}
\rvert\phi_0\rangle &= \rvert0\rangle^{\otimes N/2} \otimes \rvert1\rangle^{\otimes N/2}\\
\rvert\phi_1\rangle &= \left(\rvert0\rangle\otimes\rvert1\rangle\right)^{\otimes N/2} \;,
\end{split}
\end{equation}
in the first layer of nearest-neighbor Heisenberg interaction encoded in the unitary $W$ from Eq.~\eqref{eq:wunitary}, only the central qubits will evolve for the state $\rvert\phi_0\rangle$ while all of them will be affected for the state $\rvert\phi_1\rangle$. The full Trotter step $\mathcal{U}^{(k)}_T(t)$ from Eq.~\eqref{eq:trotter} is invariant for relabeling of qubits to neutrinos (at least for short times $t$ in general) so the final evolution is the same for both states. 

Our strategy to compress the time-evolution unitary, while keeping the average evolution of one-body observables correct, relies on randomizing the coefficients in the individual layers where $W$ acts and looking at the resulting dynamics for a varying number of layers which is in general now independent on the system size $N$. The resulting unitary takes then the following form
\begin{equation}
\label{eq:random_u}
\mathcal{U}^{(k)}_R(\tau,L) = \prod_{l=1}^{L} W(\tau,\widetilde{\bf h}^{(k,l)},{\bf 1})\;,
\end{equation}
where the components of the coupling constant vectors $\widetilde{\bf h}^{(k,l)}$ are now chosen uniformly at random from the entire $N(N-1)/2$ components of the $\mu^{(k)}_{i,j}$ coupling matrix.

For a given choice of step size, the time parameter $\tau$ is then connected to the actual total time evolution but these two are not identical. 

Our rationale for decoupling the number of layers in the evolution unitary $\mathcal{U}^{(k)}_R(\tau, L)$ from the system size $N$ is that, on average, each qubit undergoes $2L$ random pairwise interactions across the $L$ alternating even–odd layers. 

Within the Trotterized dynamics, one application of $\mathcal{U}^{(k)}_T(t)$ involves $N/2$ sequential layers, each implementing pairwise rotations with angles of order $\mathcal{O}(\mu t / N)$. For systems described by initial states with a constant fraction of orthogonal flavor state as a function of $N$, the cumulative effect of these layers produces an overall flavor rotation of order $\mathcal{O}(\mu t)$. 

Since the randomized protocol implemented using $\mathcal{U}^{(k)}_R(\tau,L)$ is designed to approximate the average long-time dynamics, in the limit of large $L$ we expect to be able to reproduce the overall rotation angle $\mathcal{O}(\mu t)$ using $L$ layers of rotation of magnitude $\mathcal{O}(\mu \tau / N)=\mathcal{O}(\mu t / L)$ each. In other words, we expect a physical time evolution for time $t$ will correspond to a value of the time-parameter $\tau$ given by
\begin{equation}
\tau = \alpha t \quad\text{with}\quad \alpha = \mathcal{O}\left(\frac{N}{L}\right)\;.
\end{equation}
This argument however neglects the effect coming from the fact that, as the partial swaps proceed, the qubits progressively rotate away from their original polarization reducing the effect of the partial swaps. Indeed for two neutrinos with the same flavor polarization, the partial swap acts simply as the identity. We therefore expect that
the total accumulated rotation angle will grow slower than linearly in the depth and we expect that $\alpha<N/L$ in practice. The actual dependence of $\alpha$ from $N$ and $L$ also strongly depends on the initial state used in the simulation as discussed previously when we considered the initial states in Eq.~\eqref{eq:init_states}. In our simulations we account for the dependence on the initial state by performing, for each individual run, a random permutation to the qubits in the state $\rvert\phi_0\rangle$ obtaining a state $\rvert\phi_s\rangle$. Physical observable are then recovered by looking at the distribution of outcomes generated by sampling both the Hamiltonian  $\hat{H}_{\nu\nu}^{(k)}$ as well as the initial state.

\begin{figure}
    \includegraphics[width=8cm]{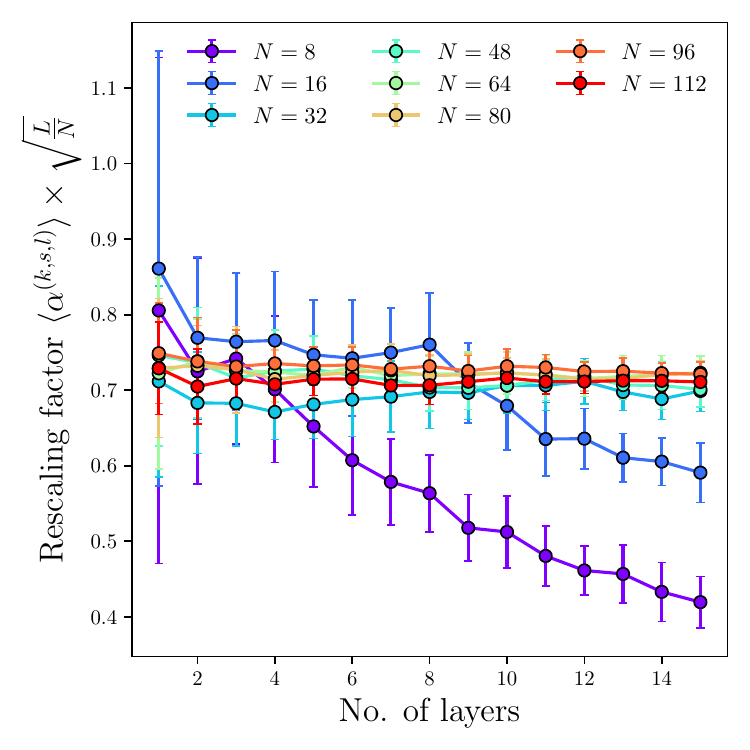}
    \caption{\justifying \textbf{Rescaling factor.} The rescaling factor normalized with $\sqrt{N/l}$  shown as a function of the number of layers for different system sizes. The expectation value and standard deviation is computed across Hamiltonian realization and initial state $(k,s)$ at fixed depth $l$. Error bars correspond to a 25\%-75\% confidence interval across Hamiltonian realizations.}
    \label{fig:rescaling}
\end{figure}

For a given initial state and Hamiltonian we then evaluate the flavor variance as
\begin{equation}
V_R^{(k,s)}(\tau,L)= \frac{1}{N}\sum_j \left(\langle\phi_s \lvert\mathcal{U}^{(k)}_R(\tau,L)^\dagger Z_j\mathcal{U}^{(k)}_R(\tau,L)\rvert\phi_s\rangle\right)^2\;.
\end{equation}
The matching between the discrete values $V_R^{(k,s)}(\tau,L)$ to the corresponding variance at a physical time $t$ is obtained through the proportionality constant $\alpha$. In order to find this factor we impose that for short times the randomized expectation value 
\begin{equation}
V_R^{(k,s)}\left(\tau,L\right)= 1+g^{(k,s,L)}_R\frac{\tau^2}{2} +\mathcal{O}(\tau^3)
\end{equation}
matches the ideal variance 
\begin{equation}
V^{(k)}(t) = 1+g^{(k)}\frac{t^2}{2}+\mathcal{O}(t^3)\;,
\end{equation}
where the linear term is zero by construction for our initial states while
\begin{equation}
\begin{split}    
    g^{(k)} &\equiv \left.\frac{d^2}{dt^2} V^{(k)}(t)\right|_{t=0} \\
    &= \frac{1}{2N} \sum_{ij} \left(\mu^{(k)}_{i,j}\right)^{2}\left( 1
        - \langle Z_i \rangle \langle Z_j \rangle \right),
\end{split}
\end{equation}
where the expectation values are evaluated on the initial state. Note that $g$ is invariant for permutation of qubits on the initial state and we can just use the value in $\rvert\phi_0\rangle$.
Using this strategy, the rescaling factor between $t$ and $\tau$ is then
\begin{equation}
\label{eq:scaling}
\alpha^{(k,s,L)} = \sqrt{\frac{g^{(k)}}{g^{(k,s,L)}_R}}\;.
\end{equation}
We show in Fig.~\ref{fig:rescaling} that the rescaling factor is in fact constant as we increase the number of layers. Note that this rescaling factor is a short time property of the compilation scheme and can thus be computed classically in an exact way before performing the simulations.

We now explore two separate protocols to perform this short time matching which are equivalent in the asymptotics but behave differently in the intermediate regime of interest. 
While our protocols share similarities in spirit with the qDrift protocol \cite{QDrift,qDRIFT_Kiss,QDRift_caltech}, the key distinction lies in the channel distance we are interested in minimizing. The goal of the qDrift protocol is to implement a random unitary channel $\Lambda_{qDrift}(t)[\rho]=\mathcal{V}_R(t)\rho\mathcal{V}^\dagger_R(t)$ whose averages matches the exact unitary dynamics at fixed couplings for short times
\begin{equation}
\left\|\mathbb{E}\left[\Lambda_{\text{qDrift}}(t)[\rho]\right]-\mathcal{U}^{(k)}(t)\rho\mathcal{U}^{(k)}(t)^\dagger\right\|=\mathcal{O}(t^2)\;.
\end{equation}
In our case instead we are first only interested in the distance between the following convex superposition of reduced density matrices
\begin{equation}
\rho_{\text{rnd}}[\rho]=\frac{1}{N!}\sum_{\sigma\in S_N}\bigotimes_{j=1}^N \text{Tr}_{N/{\sigma(j)}}[\rho]\;,
\end{equation}
where $\sigma(j)$ is an arbitrary permutation of the qubit labels taken for the symmetric group $S_N$ and $\text{Tr}_{N/{\sigma(j)}}[\rho]$ is the single qubit reduced density matrix for the $\sigma(j)$ neutrino. This is then a much weaker condition than reproducing the full density matrix itself. The distance thus reads
\begin{equation}
\label{eq:target_distance}
\left\|\rho_\text{rnd}\left[\mathcal{U}^{(k)}(t)\rho\mathcal{U}^{(k)}(t)^\dagger\right]-\mathbb{E}_{l}\left[\rho_{\text{rnd}}\left[\mathcal{U}_R^{(k)}(\tau)\rho^l\mathcal{U}_R^{(k)}(\tau)^\dagger\right]\right]\right\|\;,
\end{equation}
and the average is over initial states obtained by qubit permutations, which give rise to the same final density matrix under exact evolution. The value of $\tau$ is found by matching the short time behavior and the goal is for the randomized procedure to provide a small value for this distance, at least one averaged over the random set of Hamiltonians $\hat{H}_{\nu\nu}^{(k)}$, for times comparable to the thermalization time.

We will show below numerical evidence suggesting that our choices for implementing the randomized channel indeed are able to keep the distance in Eq.~\eqref{eq:target_distance} small. An important extension of the present work would be to provide more rigorous bounds on it.

\subsubsection{Protocol A}
One possibility is to pick a fixed value of $L$ for all times $t$ and consider

\begin{equation}
V_A^{(k,s,L)}(t) = V_R^{(k,s)}\left(\frac{t}{L}\alpha^{(k,s,L)},L\right)\;.
\end{equation}
This approach resembles a Trotter decomposition with a fixed depth where different evolution times are obtained by changing the time-step $t/L$ used for each individual layer. In the limit $L\to\infty$ the randomized evolution approximates progressively well the exact evolution under the average Hamiltonian. To see this we can look for the effective generator $H_{\text{eff}}$ by expanding at small times since we can write
\begin{equation}
\mathcal{U}^{(k)}_R\left(\frac{t}{L},L\right)=\prod_{m=1}^M\mathcal{U}^{(k)}_R\left(\frac{t}{L},\frac{L}{M}\right)\;,
\end{equation}
and for $L\to\infty$ at fixed $M$ we have 
\begin{equation}
\begin{split}
\mathcal{U}^{(k)}_R\left(\frac{t}{L},\frac{L}{M}\right)&=\prod_{l=1}^{2L/(NM)}\prod_{n=1}^{N/2} W\left(\frac{\tau}{L},\widetilde{\bf h}^{(k,Nl/2+n)},{\bf 1}\right)\\
&\approx \mathbb{1}-i\frac{t}{L}\sum_{l=1}^{2L/(NM)}\sum_{i<j}^N\mu^{(k,l)}_{i,j} P_{i,j}\\%
\end{split}
\end{equation}

so that we have as effective Hamiltonian,
\begin{equation}
\label{eq:H_eff}
H_{\text{eff}} =\sum_{i<j}^N\left(\sum_{l=1}^{2L/(NM)}\mu^{(k,l)}_{i,j}\right) P_{i,j}\to \frac{2L}{NM}\mathbb{E}[\hat{H}_{\nu\nu}^{(k)}]\;.
\end{equation}
This is then exactly the average Hamiltonian and we have
\begin{equation}
\mathcal{U}^{(k)}_R\left(\frac{t}{L},L\right)\to \exp(-i\frac{2t}{N}\mathbb{E}[\hat{H}_{\nu\nu}^{(k)}])\;,
\end{equation}
for a sufficiently large number of layers. This provides insight into the convergence of our random channel. Yet, because the dynamics under the average Hamiltonian can be completely characterized analytically and is in general different from the correct dynamical evolution, we must modify the protocol to capture the typical evolution of a random Hamiltonian.
\subsubsection{Protocol B}
This time, we instead fix a time-step $\delta t$ and vary the number of layers $L$ to reach different final evolution times.

In this way, the physical flavor variance as a function of time is estimated as
\begin{equation}
V_B^{(k,s,\tau)}(t=l\delta t) = V_R^{(k,s)}\left(\delta t \alpha^{(k,s,l)}_B,l\right)\;,
\end{equation}
for discrete times of step $\delta t$. 
The limit $L\to\infty$ now corresponds to $t\to\infty$ in steps of constant size.

\begin{figure*}
    \centering
    \includegraphics[width=18cm]{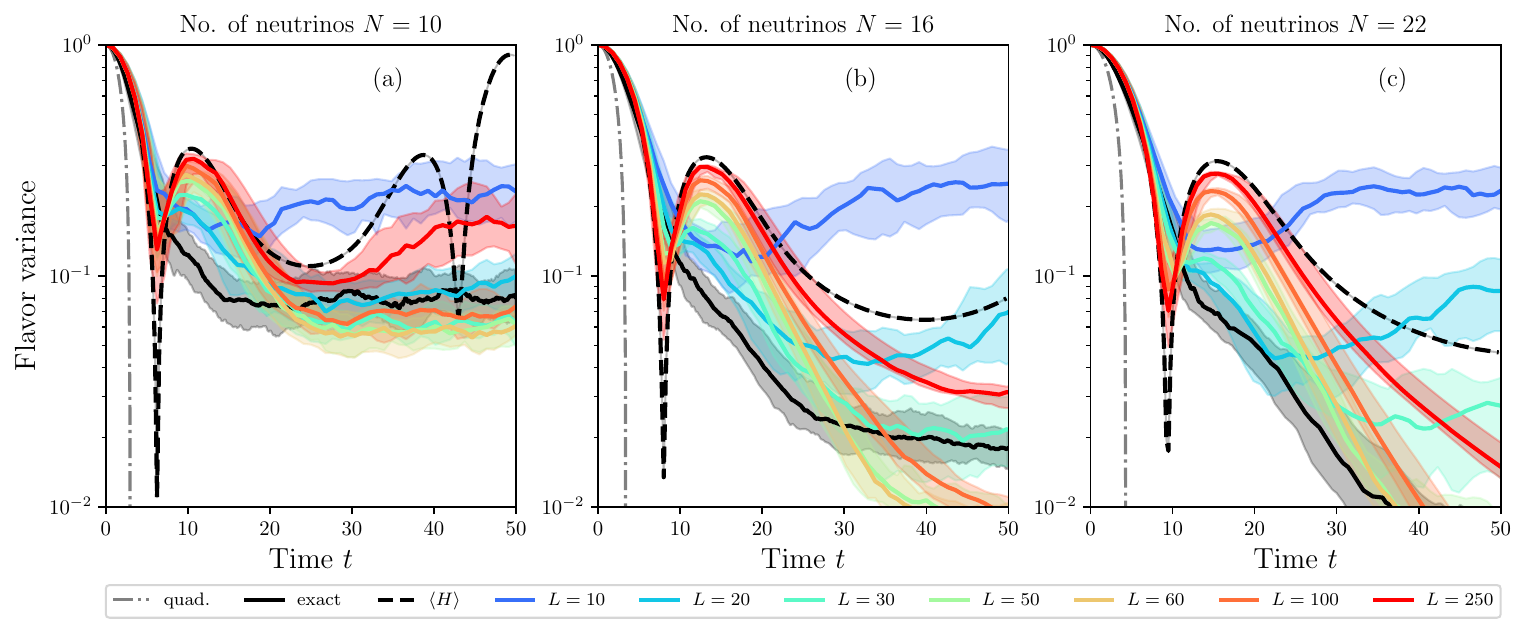}
    \caption{\justifying \textbf{Protocol A.} The variance of a single-qubit observable, denoted $V(t)$, is shown as a function of time under protocol A for different number of (fix) layers, time step $\delta t = t/L$ and system sizes: (a) $N=10$, (b) $N=16$, (c) $N=22$. The black solid line represents the exact dynamics computed using a converged product formula (time step of $0.1$), while the colored line corresponds to the evolution with the randomized protocol. The shaded band indicates a 25\%-75\% confidence interval across Hamiltonian realizations. The gray dash–dotted (“quad”) line represents the second-order Taylor expansion around $t=0$, obtained using analytical derivatives.
    }
    \label{fig:statevector-A}
\end{figure*}

\begin{figure*}
    \centering
    \includegraphics[width=18cm]{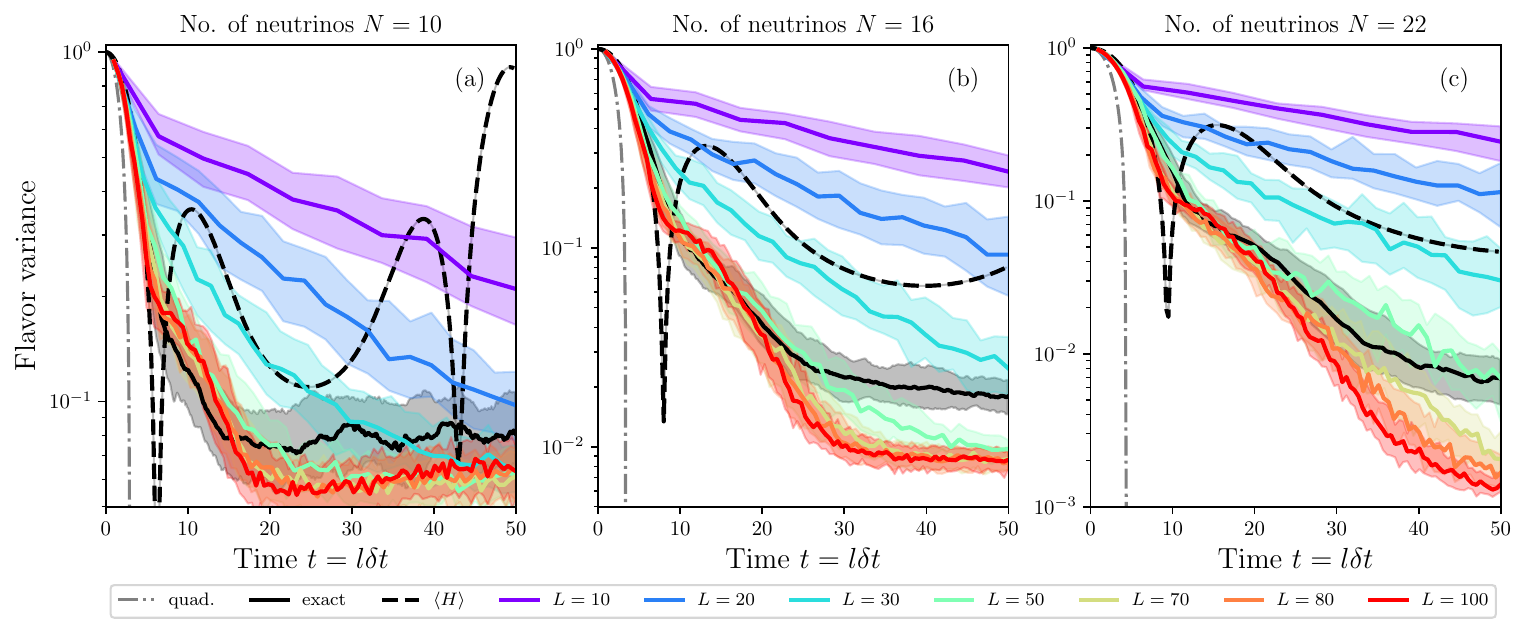}
    \caption{\justifying\textbf{Protocol B.} The variance of a single-qubit observable, denoted $V(t)$, is shown as a function of time under protocol B for different number of (fix) layers, time step $\delta t = t_\text{end}/L$ and system sizes: (a) $N=10$, (b) $N=16$, (c) $N=22$. The black solid line represents the exact dynamics computed using a converged product formula (time step of $0.1$), while the colored line corresponds to the evolution with the randomized protocol. The shaded band indicates a 25\%-75\% confidence interval across Hamiltonian realizations. The gray dash–dotted (“quad”) line represents the second-order Taylor expansion around $t=0$, obtained using analytical derivatives.
}
    \label{fig:statevector-B}
\end{figure*}

\subsubsection{Comparison between the two protocols}

We aim to compute the variance of the neutrino flavor for a typical Hamiltonian $\hat{H} \sim H$. To this end, we repeatedly sample a Hamiltonian $\hat{H}$ together with an initial state (a randomly shuffled wall configuration) and evaluate the median and 25\%-75\% percentiles  of the resulting dynamics.

We start by analyzing protocol A, shown in Fig.~\ref{fig:statevector-A}. The solid black line corresponds to the standard first-order Trotter product formula with a time step $\delta t = 0.1$. The gray dash–dotted (“quad”) line represents the second-order Taylor expansion around $t=0$, obtained using analytical derivatives. We observe that the results converge, as the number of layers increases (colors in the figure), toward the dynamics governed by the time-averaged Hamiltonian $\bar{H} = \mathbb{E}[\hat{H}_{\nu\nu}^{(k)}]$ (black dashed line). Moreover, the protocol remains accurate even beyond the validity of the second-order Taylor expansion (grey dashed line). 
However, protocol A reproduces the dynamics of a typical Hamiltonian only at short times. 

We next consider protocol B, shown in Fig.~\ref{fig:statevector-B}. Here, the dynamics remains close to the exact evolution for a time significantly longer than that predicted by the second-order Taylor expansion (grey dashed line). We observe an optimal regime around $L \simeq t_{\text{end}}$, in which the protocol maintains accuracy over the longest time window. This indicates that protocol~B provides a reliable approximation to the dynamics of a typical Hamiltonian in an intermediate-time regime.
The behavior observed for the averaged Hamiltonian at $N=10$ is attributed to finite-size effects.

For all the results discussed here, and for the remainder of the paper, we directly report physical times by using the computed rescaling factor $\alpha$ to move between $\tau$ and $t$.

\subsubsection{Computing the rescaling factor}
\label{app:variance}
Below, we provide additional details about the calculation of the short time approximation for the variance. We are interested in performing a Taylor expansion 
\begin{equation}
    V(t) = 1 + \left.\frac{d}{d t}V(t) \right|_{t=0}t +  \left.\frac{d^2}{dt^2} V(t)\right|_{t=0}\frac{t^2}{2} + \mathcal{O}(\tau^3)\;.
\end{equation}
We can compute the first derivatives as follows
\begin{equation}
\begin{split}    
   \left.\frac{d}{d t} V(t)\right|_{t=0} &=   \left.\frac{d}{d t}  \frac{1}{N} \sum_{j=0}^{N-1}
    \langle  Z_j(t) \rangle^2\right|_{t=0}  \\
    &= -\frac{2i}{N} \sum_{j=0}^{N-1} \langle Z_j \rangle \langle [H,Z_j] \rangle,
\end{split}
\end{equation}
where the commutator can be expanded as 
\begin{equation}
    \begin{split}
        [H,Z_k]  &\\
        =&\sum_{i<j} \left( \nu_{ij} [X_iX_j,Z_k] + \nu_{ij} [Y_iY_j,Z_k] + \nu_{ij} [Z_iZ_j,Z_k]\right) \\
        =& i \sum_{i<j}  \nu_{ij} \left[\delta_{ik}\left( X_iY_j -Y_iX_j\right) +  \delta_{jk} \left( Y_iX_j -X_iY_j\right)\right] \;.
    \end{split}
\end{equation}

The expectation value of this operator on a product state in the $Z$ basis is zero.

For the second derivative we instead obtain \cite{Equilibirum_neutrio_roggero}
\begin{equation}
\begin{split}
&\langle\phi_0\lvert[H, [H, Z_k]]\rvert\phi_0\rangle  \\
&= 4m_k\sum_{j,m_j=-m_k}\mu^2_{kj} \\
&= 2m_k\sum_{j=0}^{N-1}\mu^2_{kj}(1-m_jm_k)\\
\end{split}
\end{equation}
where $\rvert\phi_0\rangle = \rvert0\rangle^{\otimes N/2} \otimes \rvert1\rangle^{\otimes N/2}$ is the state at $t=0$, the factors $m_i$ are the eigenvalues of $Z_i$ on the initial state
\begin{equation}
Z_i\rvert\phi_0\rangle = m_i\rvert\phi_0\rangle\;,
\end{equation}
and the sum is over all spins $k$ with opposite polarization as the $k$-th. Using these result we finally get
\begin{equation}
\begin{split}    
\left.\frac{d^2}{dt^2} V(t)\right|_{t=0}&=-\frac{2}{N}\sum_{j=0}^{N-1}
    \langle  Z_j \rangle\langle [H,[H,Z_j]]\rangle\\
    &=-\frac{4}{N}\sum_{j=0}^{N-1} \sum_{k=0}^{N-1}\mu^2_{jk}(1-m_jm_k)\;,
\end{split}
\end{equation}
where we used the symmetry $\mu_{ij}=\mu_{ji}$ of the coupling matrix. For the specific initial sate $\rvert\phi_0\rangle$ we use here this simplifies to
\begin{equation}
\begin{split}    
\left.\frac{d^2}{dt^2} V(t)\right|_{t=0}    &=-\frac{8}{N}\sum_{j=0}^{N-1} m_j^2\sum_{k,m_k=-m_j}\mu^2_{jk}\\
    &=-\frac{8}{N}\left(\sum_{j=0}^{N/2-1} \sum_{k=N/2}^{N-1}\left(\mu^2_{jk}+\mu^2_{kj}\right)\right)\\
    &=-\frac{16}{N}\sum_{j=0}^{N/2-1} \sum_{k=N/2}^{N-1}\mu^2_{jk}\;,
\end{split}
\end{equation}
where we used the symmetry $\mu_{ij}=\mu_{ji}$ of the coupling matrix. For the full neutrino Hamiltonian we then expect on average that
\begin{equation}
\mathbb{E}\left[\left.\frac{d^2}{dt^2} V(t)\right|_{t=0}\right]=-\frac{4\mu^2}{N}\mathbb{E}\left[1-{\bf v}_1\cdot{\bf v}_2\right]\;,
\end{equation}
where the vectors ${\bf v}_1$ and ${\bf v}_2$ are sampled from the distribution in Eq.~\eqref{eq:dist_vel} of the main text.

When instead we apply a random channel $\mathcal{U}^{(k)}_R(\tau,L)$, we will write for small times $\tau$ that
\begin{equation}
\mathcal{U}^{(k)}_R(\tau,L)\approx\exp\left(-i\tau H_{\text{eff}}(k,L)\right)\;,
\end{equation}
where the effective Hamiltonian is given by
\begin{equation}
H_{\text{eff}}(k,L) = \sum_{l=1}^L \sum_{n=0}^{N-2} \widetilde{h}^{(k,l)}_{n} P_{i(n,l),j(n+1,l)}\;,
\end{equation}
where the indices $i(n,l)$ and $j(n+1,l)$ correspond to the neutrino labels occurring at layer $l$ for the $n$-the term in the even/odd evolutions implement in W from Eq.~\eqref{eq:wunitary} in the main text. These labels can be reconstructed efficiently once a given circuit layout has been generated by following the SWAP network and keeping track of the indices. The effective Hamiltonian $H_{\text{eff}}(k,L)$ can then be represented exactly as the generic $H$ in Eq.~\eqref{eq:nu_ham}  with
\begin{equation}
\mu_{ij} = \sum_{l=1}^L \sum_{n=0}^{N-2} \widetilde{h}^{(k,l)}_{n} \delta_{i,i(n,l)}\delta_{j,j(n,l)}\;.
\end{equation}

We will then expect that for the randomized unitary dynamics, the second derivative of the flavor variance should scale as
\begin{equation}
\left.\frac{d^2}{dt^2} V_R(t)\right|_{t=0} = \mathcal{O}\left(\frac{\mu^2L}{N^2}\right)\;,
\end{equation}
which will then lead to a rescaling factor
\begin{equation}
\alpha = \sqrt{\frac{\left.\frac{d^2}{dt^2} V(t)\right|_{t=0}}{\left.\frac{d^2}{dt^2} V_R(t)\right|_{t=0}}}=\mathcal{O}\left(\sqrt{\frac{N}{L}}\right)\;.
\end{equation}
\subsection{Error mitigation}
Error mitigation is an essential technique for harnessing the computational capabilities of noisy quantum devices. Our protocol builds on a noise renormalization approach, which has demonstrated strong performance in previous studies~\cite{teplitskiy2024,noise_estimation,Rahman22,kiss2024_moments}. The core idea is to model the noise as a depolarizing channel, estimate its parameters directly on hardware, and then invert its effect in post-processing.

We recall that a depolarizing noise channel $\mathcal{N}_p[\cdot]$, parameterized by $p$, affects the expectation value of a Pauli observable $\sigma$ under a quantum state $\rho$ as follows:
\begin{equation}
    \text{Tr}\left(\sigma \mathcal{N}_p[\rho]\right)= (1-p)  \text{Tr}\left(\sigma \rho\right).
\end{equation}
When the depolarizing parameter $p$ is known, noisy expectation values can be corrected by dividing them by the factor $(1-p)$. In our implementation, we estimate \( p \) using the all-zero initial state \( |0\rangle^{\otimes N} \equiv |\bar{0}\rangle\), which remains invariant under the Hamiltonian evolution
\begin{equation} 
  \langle\bar{0}| \TE{t}^\dagger Z_j \TE{t} |\bar{0}\rangle = 1.
\end{equation}
We can then use this initial state to estimate the survival probability $(1 - p)$ and employ it to renormalize the noisy estimates. Selecting an initial state that is invariant under the dynamics is, in general, a more robust strategy than attempting to invert the evolution using $\TE{t/2}\TE{-t/2}$. Indeed, such an inversion does not reproduce the same set of Pauli paths as the forward evolution in the target computation. As a result, the noise acts differently on the system, which constitutes a significant limitation of this approach. It is also worth noting that this method is most effective under the assumption of depolarizing noise, which may not fully capture the behavior of real quantum devices. Nevertheless, this limitation can be alleviated through Pauli twirling~\cite{Pauli_Twirling}. Measurement (readout) errors are addressed via device calibration~\cite{MEM} and by twirling the measurement channel~\cite{TREX}. 

\subsection{Phase-Space Approximation}
The Phase-Space Approximation (PSA) was originally introduced in nuclear physics 
\cite{Ayik2008,Lacroix2012,Lacroix2014} to treat many-body fermionic problems beyond the independent particle approximation, also called the mean-field approximation. Instead of considering a single mean-field evolution, consider a statistical ensemble of independent mean-field evolutions that differ from their initial conditions. In spirit, the PSA approach can be seen as a mapping 
from a quantum problem to a classical problem similar to what can be done 
with general quantum mechanical problems \cite{Gardiner2004}, except that here 
the mapping is made at the many-body level. The main hypothesis of the PSA is that (i) at initial time, a quantum system can be mapped to the initial ensemble
by imposing that at least the first and second moments of one-body observables
of the quantum systems match the statistical average over the ensemble. Eventually, higher moments can also be imposed \cite{Lacroix2024}; (ii) Each initial condition evolves in time using the mean-field, which is considered here as the most ``classical'' equation of motion for the set of interacting particles; 
(iii) Since the trajectories are supposed to be independent, interferences 
between trajectories that might affect the long-time evolution are neglected. 

Despite these assumption, the PSA turns out to significantly improve over the mean-field approach by providing in general rather accurate evolution of both one- and two-body observables evolutions for a wide range of coupling strengths. It was successfully applied to a wide range of physical systems ranging from permutation invariant systems presenting a quantum phase transition \cite{Lacroix2012, Lacroix2013,Lacroix2014b}, small superfluid systems, or electronic systems on a lattice \cite{Lacroix2014}. 

As shown recently \cite{Lacroix2022}, this technique offers a competitive approach to interacting-neutrino dynamics under both time-independent and time-dependent couplings.
Most recent developments of the PSA include the possibility to treat an arbitrary number of neutrino beams \cite{Lacroix2024}, assuming that each neutrinos have only two flavors, or the more realistic case where neutrinos can have three flavors \cite{Mangin-Brinet2025}. 
One key aspect of the PSA is that it inherits the simplicity of the mean-field approach, and the number of coupled equations to solve along each trajectory scales linearly with the number of neutrinos $N$ 
(as $3N$ and $8N$ for the 2- and 3-flavor case). Since the trajectories 
are independent, it can also be easily parallelized on a CPU farm. 
In Refs. 
\cite{Lacroix2024} and \cite{Mangin-Brinet2025}, the approach was shown to apply to simulations of several hundred interacting neutrinos.    

Here we use the PSA approach employing directly the technique developed in Ref. \cite{Lacroix2024} for multi-beams, with the specific limit of one neutrino per beam.
All technical details are already given in \cite{Lacroix2024} and here we only give the results obtained using the random Hamiltonian proposed in this article. 
Focusing first on the Hamiltonian (\ref{eq:hamden}), for $N$ neutrinos having two possible flavor states, the PSA evolution along each path reduces to a set of $3N$ coupled equation of motion given by:
\begin{eqnarray}
\partial_t  \vec \Sigma^{\lambda}_i(t) = \sum_{j \neq i}^{N} \mu_{ij} \vec \Sigma^{\lambda}_j  (t) \wedge \vec \Sigma^\lambda_i(t),  \label{eq:eom-psa}
\end{eqnarray} 
 where $\vec \Sigma^{\lambda}_i(t)$ is associated to the neutrino $i$ and has three real components  $\left[ X^\lambda_i (t) , Y^\lambda_i (t), Z^\lambda_i (t)\right]$, where $\lambda=1, \cdots , N_{\rm evt}$ labels the events. Denoting generically by 
${\cal P}_i$ one of the Pauli matrices associated to the neutrino ``$i$'', within the phase-space approach any quantum expectation value of a Pauli string is replaced by a statistical average over the events, i.e., 
\begin{eqnarray}
    \langle {\cal P}_{i_1} \cdots {\cal P}_{i_k} \rangle_t 
\xleftrightarrow[\text{}]{PSA} \overline{{\cal P}^{\lambda}_{i_1}(t) \cdots {\cal P}^{\lambda}_{i_k}(t) }. \nonumber 
\end{eqnarray}
${\cal P}_i$ in the left side denotes one of the Pauli operators, while  ${\cal P}^{\lambda}_{i}(t)$ denotes one of the $\vec \Sigma^{\lambda}_i(t)$
components. The average on the right-hand side of a generic observable $A$, denoted by $\overline{A^\lambda}$ stands for the statistical average over events, e.g. 
$\overline{A^{\lambda}} = (N_{\rm evt})^{-1} \sum_{\lambda} A^{\lambda}$. The PSA trajectories given by \eqref{eq:eom-psa} are solved assuming randomized 
initial conditions, such that the statistical average over events at initial time, reproduces at least the mean one-body observables and their fluctuations.
The sampling method we used assumes bi-valued probabilities as prescribed in Ref. \cite{Lacroix2024}. With this method, average evolution of one-neutrino or 
two-neutrinos observables, as well as associated one- and two-neutrinos entanglement entropies have been evaluated in general with good accuracy compared to available exact results (see \cite{Lacroix2022,Lacroix2024}). In the present work, we use the PSA method as a reference for classical simulation of large scale systems. Even if PSA does not have guarantees, it serves as an empirical independent benchmark. PSA is computationally efficient: all simulations reported here required only a few minutes of wall-clock time. Furthermore, prior results demonstrate scalability up to 2000 qubits \cite{de2026emulation}.

\section{Results and Discussion}
We are interested in the scaling of the thermalization time in systems of forward-scattering neutrinos. While a rigorous lower bound of $\log N$ can be established from the Lieb–Robinson velocity \cite{lieb1972finite}, a meaningful upper bound remains unknown. To probe thermalization, we define a practical proxy: the time required for the flavor variance  (see Eq.~\eqref{Eq_V_k_t})  to reach a threshold value $\Gamma = 0.1$. This criterion is analogous to the use of single-qubit entropy as a thermalization indicator in Ref.~\cite{Equilibirum_neutrio_roggero}.

In what follows, we systematically study this proxy for thermalization as a function of system size. Our analysis is carried out using the randomized protocol B with $L = t$, implemented across multiple computational backends: tensor-network simulations \cite{MPS_perez_garcia}, Pauli propagation methods \cite{rudolph2025paulipropagation}, and quantum processing units (QPUs) provided by IBM. This multi-platform approach enables us to benchmark the protocol’s performance, compare classical and quantum resources, and gain insight into the scaling of thermalization in regimes beyond purely analytic control. We also compare our results with the semi-classical PSA. Unlike the other methods considered, PSA does not simulate the randomized protocol directly and therefore provides an independent benchmark.

\begin{figure*}
    \includegraphics[width=17cm]{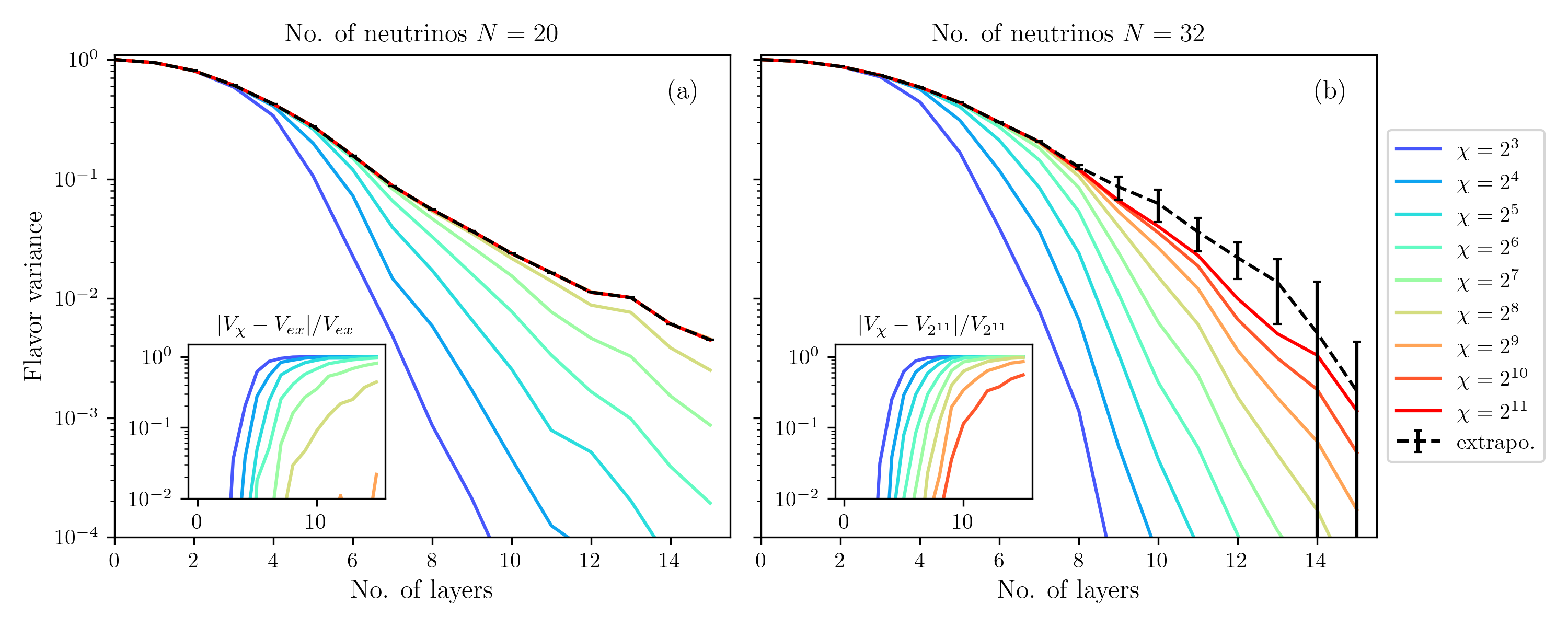}
    \caption{\justifying \textbf{MPS simulations.} The variance $V(t)$ is shown as a function of the number of layers using a fixed time step $\delta t = 1$ for different bond dimension and system sizes $N=20$ (a) and $N=32$ (b). The insets shows the relative error with the exact evolution, where for the larger system we use the largest bond dimension as a proxy. Error bars represent one standard deviation over the degrees of freedom of the extrapolation procedure, including the choice of initial points and the subset of data used in each fit.}
    \label{fig:mps}
\end{figure*}

\begin{figure}
    \centering
    \includegraphics[width=0.95\linewidth]{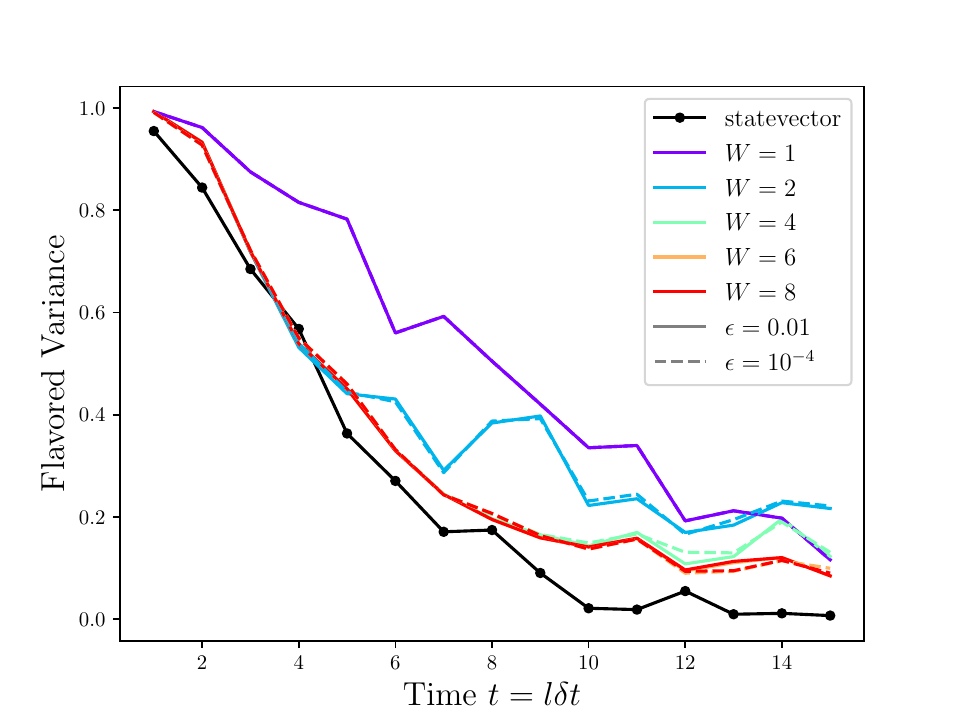}
    \caption{\justifying{\textbf{Pauli Propagation} for different $W\in [1,2,4,6,8]$ and $\epsilon \in [10^{-2}, 10^{-4}]$ for a single Hamiltonian with $N=20$ against the equivalent statevector simulation (black). } }
    \label{fig:PP-scan}
\end{figure}

\subsection{Classical Simulation}
\label{sec:classical}
We start by benchmarking three classical simulation methods. 

\subsubsection{Tensor networks} 
Tensor networks methods provide a powerful framework for simulating many-body quantum systems, particularly those with limited entanglement. Among these, matrix product states (MPS) \cite{MPS_perez_garcia} are especially effective for one-dimensional systems due to their efficient representation of quantum states with area-law entanglement. In essence, MPS encode a quantum state as a sequence of tensors connected linearly, with the bond dimension $\chi$ controlling the amount of entanglement the representation can capture.

In the following, we present numerical results obtained using MPS and demonstrate that this method is not suitable for our task, as the entanglement entropy grows linearly with time. We simulate our random channel using the \textit{quimb} library \cite{gray2018quimb}, varying the bond dimension $\chi \in 2^{\{3,\dots,11\}}$, and show in Fig.~\ref{fig:mps} that convergence requires saturating the bond dimension to $\chi = 2^{N/2 - 1}$. The inset instead shows the relative error between the MPS simulation and the exact one, where for the larger system case we take the calculation with the largest bond dimension as a proxy.

To partially mitigate this limitation, one may attempt to extrapolate observables to the formal limit $\chi \to \infty$  in order to approximate the exact result.  Treating the bond dimension as a parameter is conceptually analogous to finite-entanglement scaling approaches, where finite $\chi$ induces an effective cutoff that can be analyzed systematically~\cite{extrapolation_MPS}. In this spirit, we fit the variance $V(t;\chi)$ as a function of the bond dimension using the stretched exponential form
\begin{equation}
f(\chi) = a \exp\!\left(-b \chi^{-c}\right), 
\qquad a,b,c>0,
\label{eq:extrapolation_mps}
\end{equation}
and take the $\chi \to \infty$ limit of the fit as an estimate of the asymptotic value. We emphasize that, unlike scaling forms derived for specific critical phenomena~\cite{extrapolation_MPS}, the functional dependence in Eq.~\eqref{eq:extrapolation_mps} is empirical. The limited usefulness of MPS here is nonetheless clearly reflected by the observed bond-dimension requirements: at fixed target accuracy, convergence demands bond dimensions that grow exponentially with system size and, for the larger instances considered, approach the maximal Schmidt rank across the central bipartition, $\chi \sim 2^{N/2-1}$ (Fig.~\ref{fig:mps}). To improve robustness, we perform the fit over multiple random initial conditions and report the mean and standard deviation of the extrapolated values (black dotted line in Fig.~\ref{fig:mps}). As expected, the uncertainty increases with evolution time as entanglement builds up and truncation effects become more pronounced. Consequently, this extrapolation procedure is reliable only for moderate system sizes and evolution times where finite-$\chi$ effects remain controlled.

\subsubsection{Pauli propagation}
Pauli propagation (PP) \cite{rudolph2025paulipropagation} is a classical simulation technique, which operates in the Heisenberg picture by propagating observables backward through the quantum circuit. The method approximates the observable by expressing it as a sum over Pauli strings and truncating the expansion by discarding terms whose Pauli weight exceeds a maximum threshold $W$, or whose coefficients fall below a fixed threshold $\varepsilon$. In order to assess the behavior of PP in this setting, we perform a small-scale study shown in Fig.~\ref{fig:PP-scan}. We consider a single Hamiltonian with \(N=20\) and compare PP results for different truncation parameters \(W \in \{1,2,4,6,8\}\) and \(\epsilon \in \{10^{-2},10^{-4}\}\) against an exact statevector simulation (black curves). We observe that the PP results appear converged already for \(W=6\) and \(\epsilon = 10^{-4}\); however, despite this apparent convergence, the residual discrepancy with respect to the statevector reference remains significant. The calculation requires an hour on a personal computer.

\begin{figure*}
\includegraphics[width=17cm]{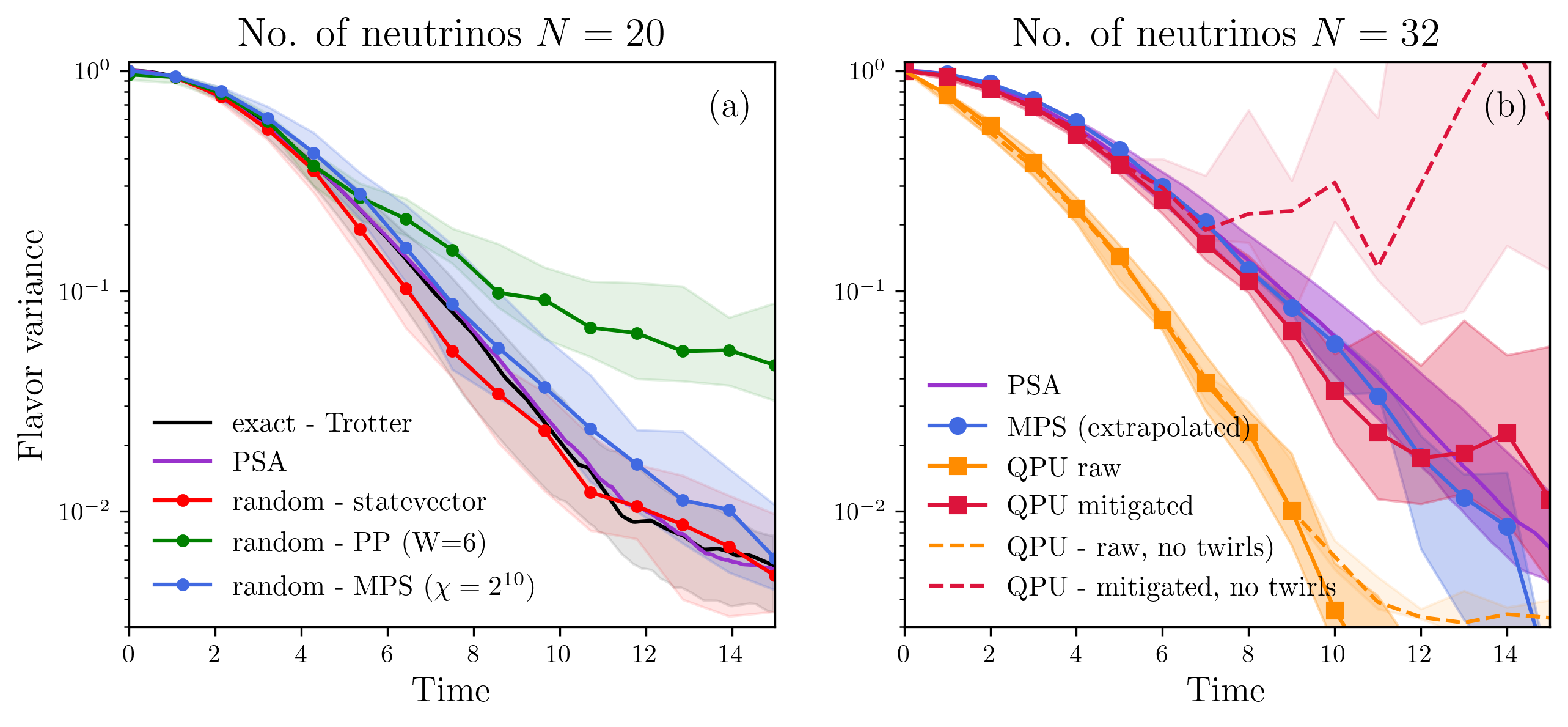}

    \caption{\justifying \textbf{Comparison between different backends.} The variance $V(t)$ is shown as a function of the number of layers using a fixed time step $\delta t = 1$ for different simulation methods. (a) we compare the exact (black), MPS (blue), Pauli Propagation (green), PSA (violet) and statevector (red) for $N=20$ while in (b) we show results on the QPU without (orange) and with (red) error mitigation compared to MPS (blue) for $N=32$. Results without twirling are shown in dashed line. The shaded band indicates a 25\%-75\% confidence interval across Hamiltonian realizations.}
    \label{fig:comparison}
\end{figure*}

\subsubsection{Comparison}
We compare all of our numerical simulation strategies in Fig.~\ref{fig:comparison}. In (a) we display the exact (black) MPS (blue), Pauli Propagation (green), PSA (violet) and statevector (red) for $N=20$ while in (b) we show results on the QPU without (orange) and with (red) error mitigation compared to MPS (blue) for $N=32$. We observe that MPS is only accurate when the bond dimension saturates the size of the Hilbert space ($\chi=2^{N/2}$), whereas PP captures only short-time dynamics and would require higher-order terms to achieve useful accuracy in this setting. On the other hand, PSA is able to reproduce the exact evolution up to long simulation times, and is thus expected to be a strong benchmark when scaling up to larger systems. Results without twirling are shown as dashed lines. We observe that it is essential to use in conjunction with our error mitigation protocol.

\subsection{Quantum Simulation}
\label{sec:ibmq}

We start by validating our error mitigation protocol on a smaller system with $N = 32$, as shown Fig.~\ref{fig:comparison} (b). This choice ensures that the MPS calculation is effectively exact, 
as, at this size, we can work at the maximal bond dimension. 
We find that our results successfully reproduce the MPS benchmark within the error bars, except for the last few points where accumulated noise becomes dominant.

\begin{figure}
    \includegraphics[width=8cm]{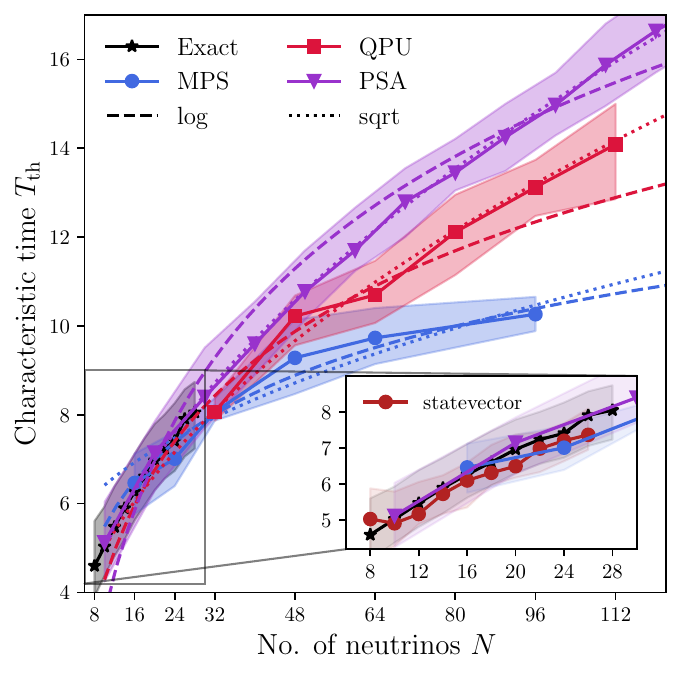}
    \caption{\justifying \textbf{Scaling of the thermalization time.} The characteristic time $T_{\text{th}}$ is shown as a function of the system size $N$ for the different methods (exact in black, MPS in blue, QPU in red, and PSA in violet), alongside square-root (dotted) and logarithmic (dashed) fits. The inset shows a zoom-in on the small system sizes (using statevector simulation instead of QPU to conserve resources). The shaded band indicates a 25\%-75\% confidence interval across Hamiltonian realizations.}
    \label{fig:thermalization}
\end{figure}

Finally, we compute the characteristic time of the variance for $N \in \{32,48,64,80,96,112 \}$ on the 156 qubits ibm devices ibm$\_$fez and ibm$\_$aachen, with typical mean two-qubit gate fidelities in the range $[99.83\%, 99.94\%]$, using the experimental setting described in Tab.~\ref{tab:experimental-settings}. The characteristic time is shown as a function of the system size in Fig.~\ref{fig:thermalization} for the different methods (exact in black, MPS in blue, QPU in red, and PSA in orange), alongside square-root (dotted) and logarithm (dashed) fits. We observe that the PSA and QPU results are compatible with each other within errorbars and exhibit a square-root scaling. For small system sizes ($N < 32$), where exact classical simulations are feasible, all methods (QPU, PSA, and MPS) show excellent agreement. For larger $N$, the MPS results start to deviate and systematically underestimate $T_{\text{th}}$, highlighting the difficulty of accurately simulating quantum dynamics. For large system sizes, both the QPU and PSA results continue to increase, exhibiting a scaling behavior compatible with $\sqrt{N}$. The PSA yields slightly higher values of $T_{\text{th}}$ than the QPU. It remains unclear whether this discrepancy originates from the approximate nature of the PSA or from residual errors in the QPU results despite mitigation. As indicated in Fig.~\ref{fig:comparison} (b), unmitigated evolutions (orange curve) tend to underestimate $T_{\text{th}}$, suggesting that imperfect error mitigation could account for the observed difference. 

We can nevertheless conclude from these results that the scaling of the thermalization time obtained using both PSA and our randomized methods is consistent, as the square root fit is the best in both cases.

\section{Conclusions}

We have introduced a randomized quantum approach capable of efficiently capturing global properties in typical random systems. Although we do not provide errors guarantees, our numerical results for small systems show that the method indeed reproduces on average the exact flavor dynamics for single qubit observables while requiring only a $\mathcal{O}(t)$ maximal circuit depth. Moreover, we demonstrate that this class of problems is challenging for tensor-network methods such as MPS, due to the linear growth of entanglement with time.

Applying our approach to systems of neutrinos interacting only through forward-scattering we find that the thermalization time scales as the square root of the system size. This behavior is consistent with the the semi-classical PSA approach and thus providing strong supporting evidence for the validity of the semi-classical treatment of this problem. This result illustrates how NISQ devices can serve as empirical validators for classical approximation methods. Thanks to their better scaling with system size, these semi-classical methods can then be used to both explore larger systems or to tackle more challenging simulation problems. In the context of flavor evolution of neutrinos this includes both the extension to the full three flavor case~\cite{PhysRevD.111.063038,gjr1-lf8s,PhysRevD.111.043038,PhysRevD.111.043017,manginbrinet2025threeflavorneutrinooscillationsusing} and the inclusion of non-forward scattering between neutrinos~\cite{PhysRevD.110.123028}.

More broadly, our work demonstrates an application of random quantum circuits (RQC) in physics that has so far received limited attention. While RQCs are known to be difficult to simulate classically~\cite{quantum_supremacy}, their use in physical modeling has remained limited. Here, we show how RQCs can provide physical insight by connecting them to the dynamics of random Hamiltonians.

Finally, our work falls into so-called quantum utility experiments: where NISQ devices are used as experimental platforms to test, benchmark, and refine empirical classical methods. Although our current implementation is limited to specific observables and lacks rigorous guarantees, it opens the door to extending such quantum-assisted benchmarking to more complex quantities, such as two-point correlation functions \cite{two_point_roggero}.

\begin{table}[h]
\centering
\caption{Experimental settings.}
\begin{tabular}{ll}
\hline 
\hline

No. of configurations & $10\times L$ \\
No. of twirls (total)         & 75 \\
No. of shots           & 500 \\
Quantum devices used      & \texttt{ibm\_fez}, \texttt{ibm\_aachen} \\
Maximum number of layers  & $L=15$ \\
Time step & $\delta t = 1$ \\
CNOT depth                & $3 \times 2 \times L\leq 90 $\\
Error mitigation overhead & $\times2$ \\
QPU time & 22-24 [s] per config. \\
\hline 
\hline
\end{tabular}
\label{tab:experimental-settings}
\end{table}

\section*{Data Availability}
The data generated by the quantum computer is publicly available~\cite{data}.

\section*{Code Availability}
The code supporting the findings of this work is  publicly available~\cite{data}.

\begin{acknowledgements}
We thank J. Carlson, N. Mueller and D. Neill for discussions about the topics covered in this work. We acknowledge the use of IBM Quantum services for this work. The views expressed are those of the authors and do not necessarily reflect the official policy or position of IBM or the IBM Quantum team.
\end{acknowledgements}

\section*{Author Contributions}
O.K. and A.R. conceived the study and developed the research plan. A.R. developed the theoretical framework. O.K. carried out the classical numerical simulations and the quantum hardware experiments, with contributions from I.T. and F.T. D.L. performed the PSA calculations. A.R. supervised the project. All authors discussed the results and contributed to the final manuscript.

\section*{Funding Statement}
O.K.\ is supported by CERN through the CERN Quantum Technology Initiative and the University of Geneva [Doc.\ Mobility fellowship]. D.L.\ discloses support for the research of this work from CNRS [AIQI-IN2P3 project] and the French National Research Agency [ANR-22-PNQC-0002] (HQI initiative, France 2030). I.T.\ and F.T.\ disclose support for the research of this work from the Swiss National Science Foundation [grant number 225229].

\section*{Competing Interests}
All authors declare no competing interests. 

\bibliography{bibliography}

\end{document}